# The mechanism of the decrease of barriers for oxygen ionic conductivity in nanocrystalline ceramics

M.D.Glinchuk[1], P.I.Bykov[2], B.Hilczer[3]

[1]*Institute for Problems of Materials Science, NASc of Ukraine, Kjijanovskogo 3, 03680 Kiev 142, Ukraine, dep4@materials.kiev.ua*

[2]*Radiophysics Faculty of Taras Shevchenko National University of Kiev, 2, Acad. Glushkov Ave., building # 5, 03127 Kiev, Ukraine*

[3]*Institute of Molecular Physics, M. Smoluchowskiego 17, 60-179 Poznan, Poland*

**Abstract**

We calculated the influence of surface tension on the barriers of oxygen ionic conductivity in nanograin ceramics. Activation energy of oxygen ions diffusion via oxygen vacancies which were considered as the dilatational centers was calculated. This energy was shown to decrease with nanoparticle sizes decreasing. The distribution function of activation energy was derived on the basis of distribution of nanoparticle sizes. We obtained an analytical expressions of ionic conductivity dependence on the temperature and nanograin sizes. These formulas fitted pretty good the observed earlier behaviour of oxygen conductivity in nanograin ceramics of $ZrO_2$:16% Y observed earlier. Therefore the consideration we carried out had shown that the surface tension in nanoparticles is physical mechanism responsible for the essential enhancement of the oxygen ionic conductivity observed in nanograin samples, the main contribution to the conductivity being related to the region in vicinity of the particle surface.

## 1. Introduction

In the last years nanocrystalline materials attract much attention of the scientists and engineers because of their unique physical properties with the anomalies caused by size effects. These effects are related to the contribution of surface that increase with the nanoparticle sizes decrease. Because of this the properties anomalies manifest themself mainly in the ceramics with grain sizes spatially confined to less than 100 nm. A number of studies have shown that such nanograin materials are characterized by improved optical, electrical and mechanical properties, which can lead to useful technical applications [1-3]. In particular the enhancement of electrical conductivity observed in [4,5] have created a new challenge for solid state ionic materials and can impact a number of applications such as batteries, solid oxide fuel cells, gas sensors and ionic membranes [6-8]. To obtain the materials with optimal properties and in particular with high ionic conductivity the understanding of mechanism of this conductivity enhancement in nanograin oxides seems to be extremely important. Although available experimental data have shown the increase of ionic conductivity in comparison with bulk in two-three orders of magnitude in nanocrystalline $ZrO_2$:16% Y with average grain size $\bar{R} = 20$ nm [5] there is no explanation of this phenomenon up to now.

     In present paper we propose the mechanism of ionic conductivity enhancement in nanograin ceramics related to the influence of surface tension on the activation energy of ionic diffusion. We also took into account the distribution of the nanoparticles sizes, which usually exist in real ceramics, and calculated the activation energy distribution assuming Gaussian distribution of particle sizes. The smearing of activation energy distribution function and decrease of the most probable value of activation energy with increase of the width of size distribution function was obtained. The theory fitted pretty good the temperature and size dependence of oxygen ionic conductivity in $ZrO_2$:16% Y reported in [5].

## 2. Model

The influence of surface on nanomaterials properties is known to be the reason of the properties anomalies. Surface energy is related to surface tension ε, which for the nanoparticle of spherical form with radius $R$ can be written as [9]

$$\varepsilon = \frac{2\alpha}{R} \qquad (1)$$

where α is surface tension coefficient. Taking some average probable value of this coefficient in oxides $\alpha \approx 50$ N/m [10], one can see, that in the range, where size effects are essential and detectable ($10 \leq R \leq 100$ nm), mechanical tension lay in the region $10^9 \leq \varepsilon \leq 10^{10}$ N/m$^2$. Therefore nanoparticles are under influence of strong hydrostatic pressure, that in $10^4$–$10^5$ times larger then the pressure of the atmosphere.

In what follows we will consider the densely packed nanoparticles as the model for nanograin ceramics in which any grain (particle) is under influence of the mechanical pressure.

It has been shown experimentally, that external mechanical pressure decreases the ions diffusion barriers because it induced an internal deformation, that changes the barrier [11]. One can see that origin of the internal deformation in nanomaterials can be the surface tension (see Eq. (1)). To find out how this deformation will influence the ionic conductivity let us consider the one dimensional diffusion of oxygen ions via oxygen vacancies in oxides nanograin ceramics. It is obvious that the diffusion of the vacancies in some direction is equivalent to the diffusion of oxygen ions in the inverse direction. Because of this we will consider the diffusion of oxygen vacancies. The oxygen vacancy can be considered as dilatational center with elastic dipole moment equal to the volume of the vacancy $P = -V$ [12]. This dipole moment has to "feel" the surface tension-induced deformation and so the additional energy of oxygen vacancy in nanoparticles can be written as

$$\Delta W = -\frac{2\alpha}{R}P = \frac{2\alpha V}{R} \qquad (2)$$

where $R$ can be considered as average grain size in the nanograin ceramics.

This additional energy can decrease the barrier for diffusion in one direction and increase it for diffusion in opposite direction so that the resulting flow can be the difference between them, i.e. [11]

$$K = 2K_0 sh\frac{\Delta W}{kT} \qquad (3)$$

where $K_0$ is related to thermal activation process, namely

$$K_0 = K_1 \exp\left(-\frac{E_0}{kT}\right) \qquad (4)$$

The coefficient $K_1$ depends on atomic vibration of the lattice and $E_0$ is activation energy of self-diffusion process in the bulk.

It follows from Eqs. (3), (4), that the ionic conductivity can be represented in the form

$$I = A\exp\left(-\frac{E_0}{kT}\right)sh\left(\frac{\Delta W}{kT}\right) \qquad (5)$$

The coefficient $A$ has dimension of the conductivity and its value depends on the characteristics of the sample and ions as well as concentration of the vacancies. The size dependence of ionic conductivity follows from Eq. (2), so that

$$I = A\exp\left(-\frac{E_0}{kT}\right)sh\left(\frac{2\alpha V}{kTR}\right) \qquad (6)$$

The expression for activation energy $E$ can be obtained from Eq. (6) by conventional way:

$$E = k\frac{d\ln(I/A)}{d(T^{-1})} = E_0 - \frac{2\alpha V}{R}cth\left(\frac{2\alpha V}{RkT}\right) \qquad (7)$$

One can see from Eq. (7) that the diffusion barrier in the case of nanoparticles decreases due to influence of surface tension, its value being dependent on the particles sizes. To analyse this dependence as well as physical meaning of activation energy temperature dependence, that seems unusual, let us consider the obtained results in more the details.

## 3. Ions diffusion activation energy and its distribution in nanograin ceramics

3.1. Activation energy

In Fig. 1 we represented temperature and size dependence of ion diffusion activation energy calculated on the basis of Eq. (7). One can see that for all the sizes there is the broad region where activation energy is practically independent of temperature and this region increases with the particle sizes decrease. Moreover the temperatures where $E(T)$ dependence becomes pronounced correspond to $T \geq 2000$ K which is out of the temperature region of ionic conductivity we are interested in (e.g. working temperature of solid oxide fuel cells is around 1200 K [6]). Because of this in what follows we will consider the temperatures $T \leq 1500$ K where temperature dependence of activation energy is negligibly small even for micron size particles (see curve 1 in Fig. 1). The decrease of diffusion barrier with nanoparticles sizes decrease, namely at $R < 100$ nm, is clearly seen in Fig. 1. Of course the value of barrier decrease depends on the parameters in Eq. (7) and mainly on $\alpha$ and $V$ values or more precisely on their product. We took for calculations represented in Fig. 1 the volume of $O^{2-}$ vacancy as $V$, i.e. $V = 1.68 \cdot 10^{-3}$ nm$^2$ and $\alpha = 50$ N/m as some average value of surface tension coefficient for oxides. For such the parameters activation energy for nanoparticle with $R = 10$ nm and $R = 20$ nm is about 2 and 1,5 times respectively smaller than for micron-sized particles, although for $R > 100$ nm $E$ is close to $E_0$, Note, that in Fig. 1 and in the other Figures we took $E_0 = 1,23$ eV that corresponds to ZrO$_2$:16% Y bulk material [5]. For the same parameters in the inset to Fig. 1 we depicted the dependence of activation energy on the particles size for several temperatures actual for oxygen ionic transport. Fast decrease of $E$ value for $R \leq 40$ nm is clearly seen for all the considered temperature while for 40 nm $< R <$ 100 nm $E(R)$ decreases slowly and its value becomes a little bit smaller with the temperature increase. For the sake of illustration in Fig. 2 we depicted the size dependence of activation energy for several value of surface tension coefficient $\alpha$. It is seen that although $E$ value practically coincide for all chosen $\alpha$ at $R \geq$

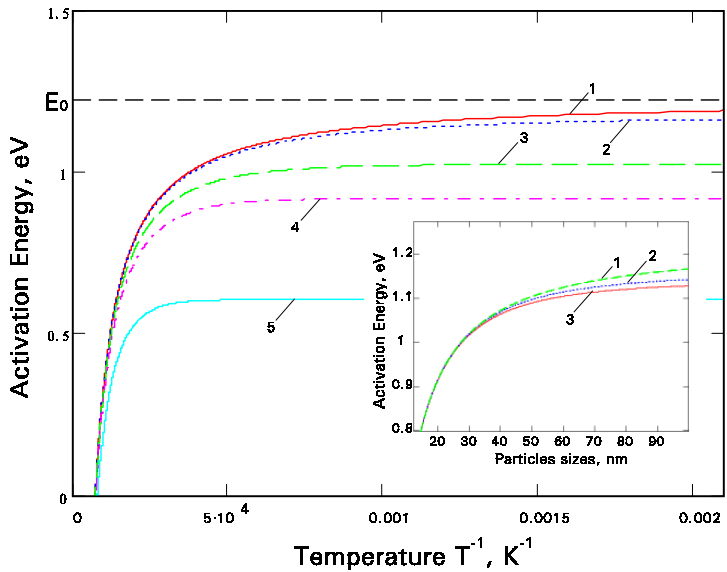

**Fig. 1.** Temperature dependence for several nanoparticle sizes $R$ (nm): 1200 (1); 100 (2); 30 (3); 20 (4); 10 (5) (basic plot) and particles size dependence for several temperatures $T$ (K): 100 (1); 800 (2); 1000 (3) (inset) of ions diffusion activation energy.

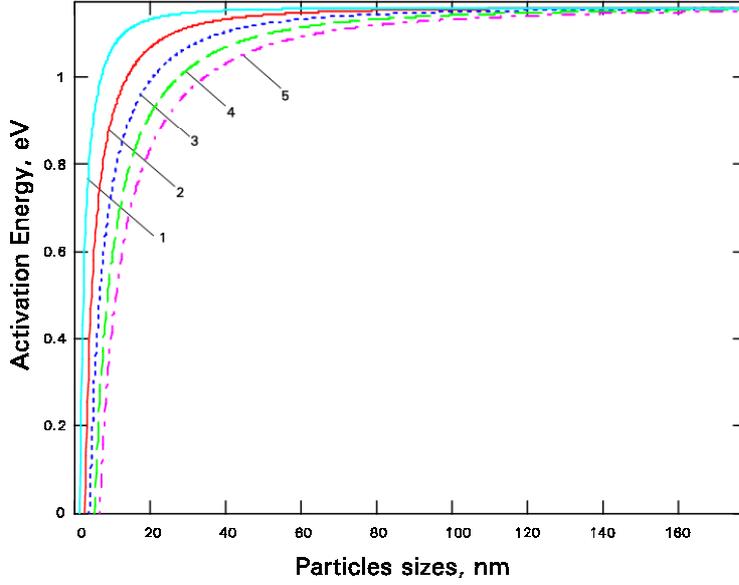

**Fig. 2.** Ions diffusion activation energy dependence particle sizes for different surface tension coefficient $\alpha$ (N/m): 25 (1); 50 (2); 75 (3); 100 (4); 125 (5) at $T = 800$ K.

100 nm, there is an essential difference between the curves at $R < 100$ nm. Therefore measurements of ion diffusion activation energy of nanograin ceramics samples with average grain sizes $R \leq 100$ nm will be used as a method of $\alpha$ measurement on the basis of comparison of observed and calculated $E(R)$ behaviour.

### 3.2. The distribution function of activation energy

In real nanograin ceramics there is the distribution of grain sizes with the parameters dependent on technology of sample preparation. This distribution can influence essentially the observed properties (see e.g. [13]) and so it has to be taken into account. Let us assume that distribution function of sizes $F(R)$ has Gaussian form, i.e.

$$F(R) = \frac{1}{\Delta\sqrt{\pi}\left(erf(R_0/\Delta)+1\right)} \exp\left(\frac{R-R_0}{\Delta}\right)^2 \qquad (8)$$

where $R_0$ and $\Delta\sqrt{\ln 2}$ are respectively the most probable size and half-width on half-height.

Average value of particles size $\bar{R}$, that can be measured experimentally, related to the distribution function parameters as

$$\frac{\bar{R}}{\Delta} = \frac{R_0}{\Delta} + \frac{\exp(-R_0/\Delta)}{\sqrt{\pi}\left(1+erf(R_0/\Delta)\right)} \qquad (9)$$

One can see that for narrow distribution function, namely at $R_0/\Delta > 1{,}5$  $R_0 = \bar{R}$ [13].

Allowing for the relation between activation energy and radius of nanoparticles given by Eq. (7) one can calculate the distribution function of activation energy with the help of statistical physics relation [14]:

$$f(E) = F(R)\left|\frac{dR}{dE}\right| \qquad (10a)$$

The calculations with respect to Eqs. (7) and (8) yields:

$$f(E) = \frac{\exp\left(-(R-R_0)/\Delta\right)^2}{\Delta\sqrt{\pi}\, erf\left(\frac{R_0}{\Delta}+1\right)\left|\frac{\alpha V}{R^2} cth\frac{\alpha V}{RkT} + \frac{\alpha^2 V^2}{R^3 kT}\left(1-cth^2\frac{\alpha V}{RkT}\right)\right|} \qquad (10b)$$

The Eq. (10b) gives the dependence of activation energy distribution function on particles sizes, surface tension coefficient and volume of oxygen vacancy, $f(E)$ temperature dependence being negligible small at $T \leq 2000$ K. Since all these factors are gathered in the expression for activation energy (see Eq. (7)) it is reasonable to depict $f(E)$ as the function of $E$ (see Fig. 3). One can see that with increase of the width of sizes distribution function the most probable activation energy increases, the distribution function shape becomes asymmetrical with broad left hand side shoulders. This behaviour is related to the fact, that activation energy increases for larger particles and decreases for smaller particles, such particles ratio being dependent on distribution function width. It is obvious that because of distribution of activation energy all physical quantities which depend on it have to be distributed also. Therefore the calculations of these quantities average values have to be performed with the help of distribution function of either sizes or activation energy.

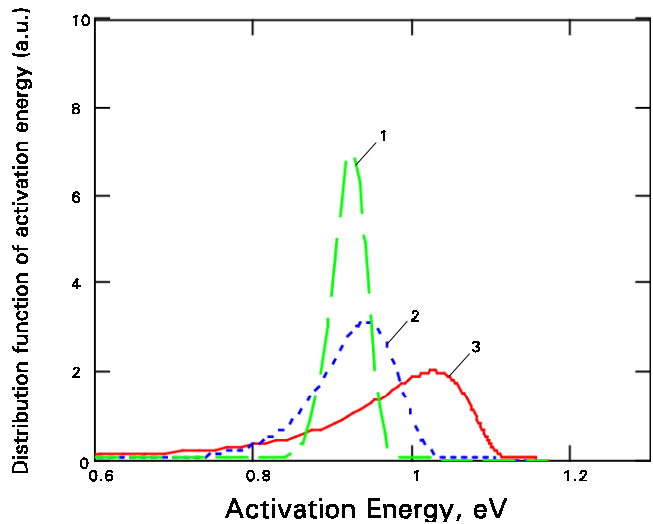

**Fig. 3.** Distribution function of ions diffusion activation energy for different nanoparticles size distribution function (see Eq. (8)) parameters Δ (nm): 2 (1); 10 (2); 20 (3) and $R_0$ = 20 nm at $\alpha$ = 50 N/m

## 4. Ionic conductivity in nanoparticles

The dimensionless ionic conductivity can be calculated on the basis of Eq. (6) as the quantity $\sigma = I/A$. One can see that it depends on temperature and particles size. The detailed form of this dependence is shown in Fig. 4. For actual temperature region $T \leq 2000$ K the conductivity increased linearly as a function of $1/T$ with temperature lowering, the slope of the straight lines related to activation energy. It is seen that slope and so activation energy decrease with the size decrease from 10 nm to 200 nm (see curves 1–3 in Fig. 4), while it practically the same for larger sizes. This behaviour is in complete agreement with the results of previous section.

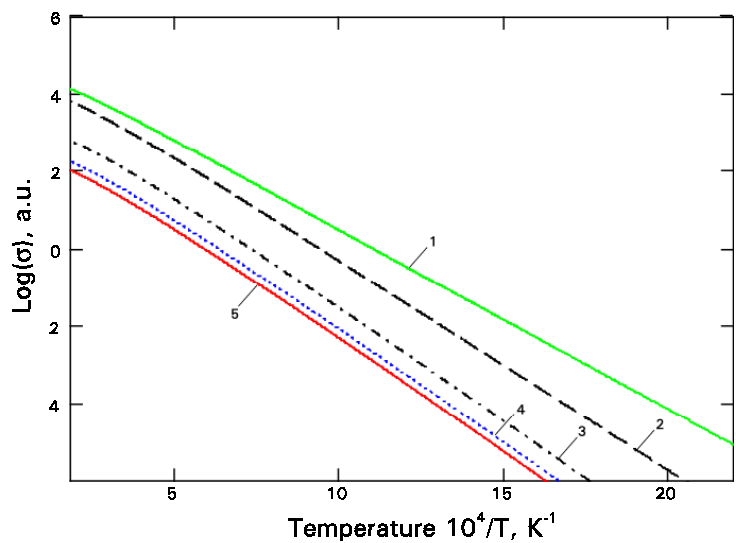

**Fig. 4.** Temperature dependence of ionic conductivity for different nanoparticles sizes $R$ (nm): 10 (1); 20 (2); 200 (3); 1000 (4); 1200 (5) at $\alpha$ = 25 N/m.

The model we used for description of ionic conductivity in nanoparticles is based on the influence of surface energy related to surface tension on the diffusion of oxygen ions via the oxygen vacancies. However it was shown by consideration of ferroelectric nanoparticles with taking into account surface tension contribution [15] that the changing of properties is especially strong in some region in vicinity of surface (shell region) while they are close to those in bulk in the central part of a particle (core region). The model of "shell", which "feels" the influence of surface and "core" with the properties like those in the bulk was proposed recently for description of radiospectroscopy spectra in the oxides [16]. Since the ratio of the shell to the core contribution increases with the particles sizes decrease the proposed model allowed to explain successfully the observed transformation of NMR spectra of $^{17}$O in MgO nanopowder from one line characteristic to bulk sample into two lines originated from core and shell with the sizes decrease. The measurements of ESR spectra of ZrO$_2$:8% Y confirmed the existence of shell and core regions, namely in nanoparticles with average size about 30 nm the size of shell was shown to be a few nm [17]. From this point of view it is not excluded that the enhancement of ionic conductivity in nanograin ceramics is related to the contribution of the nanoparticles shell region.

On the other hand namely this region can be enriched by the impurities and defects and in particular by oxygen vacancies. Allowing for the vacancies similarly to conventional electronic centers [18] will "feel" the influence of surface in its vicinity one can suppose that the largest part of the ionic conductivity enhancement can be related to the contribution of the shell regions of the particle.

To our mind the term "grain boundary" with thickness of a few nm used in the experimental works (see [5] and ref. therein) and our term "shell region" of the particle are close to one another. The contribution of intergrain space into ionic conductivity was shown to be independent on particles sizes [5] and so it can be excluded from consideration.

## 5. Comparison with experiment

The detailed measurements of ZrO$_2$:16% Y nanograin film with thickness 330 nm were carried out recently [5]. Since all the properties were shown to be practically the same as the properties of bulk polycrystal for sizes larger than 100 nm we will consider the specimens as nanograin ceramics.

Because the mechanical tension related to the surface tension is a key point of our model, let us begin with comparison of the theory with observed dependence of mechanical strain on nanoparticles size. This dependence was obtained on the basis of analysis of the shape of X-ray diffraction lines. Allowing for the linear coupling via elastic modulus $S$ between mechanical strain $\Delta d/d$ and tension $\varepsilon$ given by Eq. (1), namely $\varepsilon = S\Delta d/d = 2\alpha/\overline{R}$, the linear dependence of the strain on inverse nanograin size can be expected. It follows from Fig. 5 that this dependence (solid line in the Fig.) was really observed experimentally (filled circles) for all the specimens but one with the smallest average particles size. In our view it may be related to the increase of inaccuracy of measurements for extremely small nanograins. Note, that size independent part of the strain can be related to contribution of the particles core. Allowing for $S \cong 3 \cdot 10^{12}$ N/m$^2$ for ZrO$_2$:16% Y$_2$O$_3$ [19] it is easy to estimate surface tension coefficient value from experimental points, which are laying on the strait line in Fig. 5. This leads to $\alpha \cong 15$ N/m.

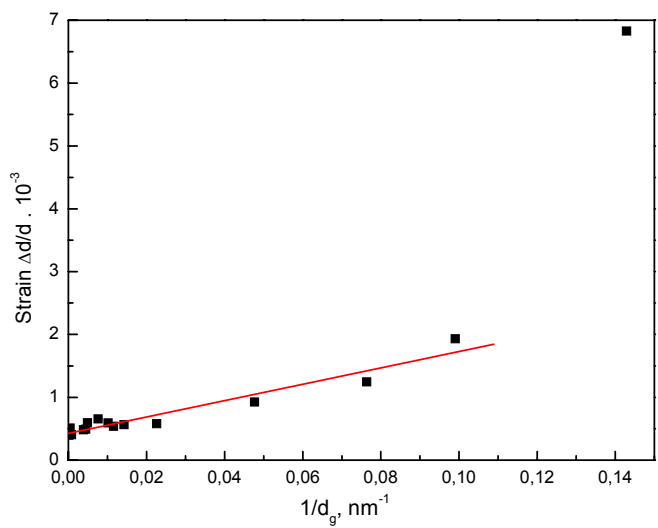

**Fig. 5.** Mechanical tension dependence on nanoparticles size. Solid line – theory (see Eq. (1)); filled circles – experiment [5]

Let us proceed to comparison of measured and calculated oxygen ionic conductivity and to the description of experimental data for $ZrO_2$:16% Y nanograin ceramics [5]. Keeping in mind the distribution of nanograin sizes in the ceramics and that only the average value of ionic conductivity can be measured, we have to average Eq. (6) with distribution function $F(R)$ of sizes given by Eq. (8), i.e.

$$\frac{I_0}{A} = \int_0^\infty F(R)\exp\left(-\frac{E_0}{kT}\right) sh\left(\frac{2\alpha V}{RkT}\right) dR \qquad (11)$$

To estimate the unknown value of surface tension coefficient $\alpha$ let us simplify the Eq. (11). In particular, allowing for that the samples $ZrO_2$:16% Y have very narrow distribution of grain sizes [5] one can rewrite Eq. (11) as

$$\frac{I_0}{A} \approx \exp\left(-\frac{E_0}{kT}\right) sh\left(\frac{2\alpha V}{\bar{R}kT}\right) \qquad (12)$$

where $\bar{R} = R_0$ because for narrow distribution average and the most probable values of particle sizes coincide (see Eq. (9)).

Fitting of the observed temperature dependence of oxygen conductivity with the theoretical expression (11) was performed at $\alpha = 25$ N/m and $\Delta = 0.45$ nm, $R_0 = 10$ nm for nanoparticles with $\bar{R} = 10$ nm, $\Delta = 65$ nm, $R_0 = 1200$ nm for the particles with $\bar{R} = 1200$ nm. One can see from Fig. 6, that the experimental points are fitted very good by solid line ($\bar{R} = 10$ nm) or dotted line ($\bar{R} = 1200$ nm) depicted with the help of Eq. (11). The parameters $\Delta$ and $R_0$ of distribution function used for fitting show that the size distribution was really narrow, so that the approximate Eq. (12) can be successfully used instead of the Eq. (11). Note that the fitting of the experimental data with Eq. (11) at $\alpha = 20$ N/m ($R_0 = 10$ nm, $\Delta = 0,475$ nm) or 15 N/m ($R_0 = 10$ nm, $\Delta = 0,525$ nm) also leads to quite good agreement of the theory with experiment, but the value $\alpha = 25$ N/m gives the best fitting. Therefore one can suppose that value of $\alpha$ for the $ZrO_2$:16% $Y_2O_3$ can be somewhere in between 15 N/m and 25 N/m. The comparison of the observed size dependence of the ionic conductivity originated from the grain boundaries with the theory is given in Fig. 7. It is seen that Eq. (12) without free parameters fitted experimental points good enough. This speaks in favour of essential contribution of the shell regions of the particles to the ionic conductivity. Therefore the consideration we carried out had shown that the surface tension in nanoparticles is the physical mechanism responsible for the essential enhancement of the oxygen ionic conductivity observed in nanograin samples, the main contribution to the conductivity being related to the shell region of the particle.

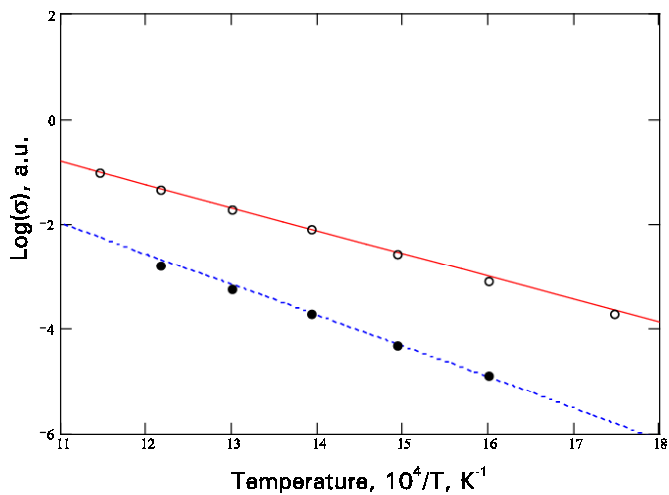

**Fig. 6.** Comparison of the theory (straight lines) with experiment for $\bar{R} = 10$ nm (open circles) and $\bar{R} = 1200$ nm (filled circles) [5] for temperature dependence of ionic conductivity.

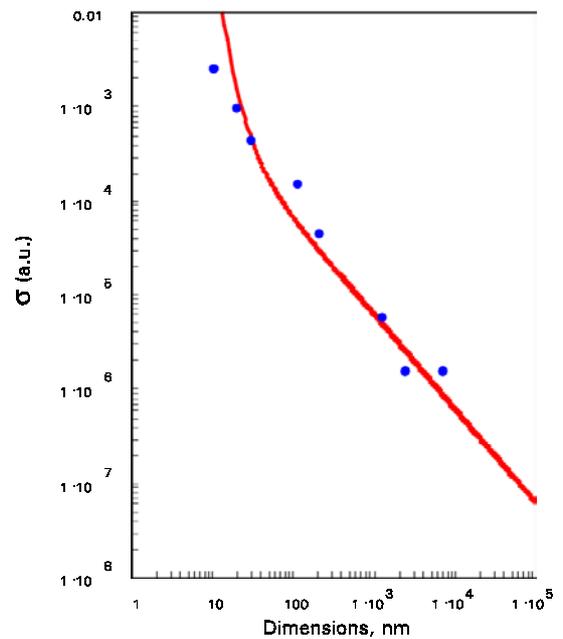

**Fig. 7.** Size dependence of ionic conductivity. Solid line – theory; circles – experiment [5].